\documentstyle[12pt,axodraw]{article}

\catcode`@=12
\newcommand{\beq}{\begin{equation}}
\newcommand{\eeq}{\end{equation}}
\newcommand{\nn}{\nonumber}
\newcommand{\bea}{\begin{eqnarray}}
\newcommand{\eea}{\end{eqnarray}}

\newcommand{\rfn}[1]{(\ref{#1})}
\newcommand{\Eq}[1]{Eq.~(\ref{#1})}

\newcommand{\epj}[1]{{ Eur. Phys. J. }{\bf #1}}

\begin{document}
\thispagestyle{empty}
\begin{flushright} UCRHEP-T296\\ December 2000\
\end{flushright}
\vspace{0.5in}
\begin{center}
{\Large	\bf Three Active and Two Sterile Neutrinos \\in an E$_6$ 
Model of Diquark Baryogenesis\\}
\vspace{1.5in}
{\bf Ernest Ma$^1$ and Martti Raidal$^{1,2}$\\}
\vspace{0.2in}
{\sl $^1$ Physics Department, University of California, Riverside, 
California 92521, USA\\}
{\sl $^2$ National Institute of Chemical Physics and Biophysics, 
Tallinn 10143, Estonia\\}
\vspace{1.1in}
\end{center}
\begin{abstract}\
In the $U(1)_N$ extension of the supersymmetric standard model with $E_6$ 
particle content, the heavy singlet superfield $N$ may decay into a quark and 
a diquark as well as an antiquark and an antidiquark, thus creating a 
baryon asymmetry of the Universe.  We show how the three doublet and two 
singlet neutrinos in this model acquire mass from physics at the TeV scale 
without the benefit of using $N$ as a heavy right-handed neutrino. 
Specifically, the active neutrinos get masses via the bilinear term $\mu LX^c$
which conserves R-parity, and via the nonzero masses of the sterile 
neutrinos.  We predict fixed properties of the extra $Z'$ boson, as well as 
the new lepton doublets $X$ and $X^c$, and the observation 
of diquark resonances at hadron colliders in this scenario. 
\end{abstract}

\newpage
\baselineskip 24pt

\section{Introduction}

There are two important issues regarding any extension beyond the minimal 
Standard Model (SM) of particle interactions.  One is the implementation 
of a natural mechanism for small Majorana neutrino masses.  This is 
highly desirable for understanding the current data on atmospheric and 
solar neutrino oscillations \cite{osc}.  The other is the implementation 
of a natural mechanism for generating a baryon asymmetry of the Universe. 
With the addition of three heavy right-handed singlet neutrinos, both 
can be achieved.  Unfortunately, this minimal extension of the SM is not 
subject to direct experimental verification at future colliders 
\cite{direct,bil}.

If there is new physics at the TeV scale, it should be such that the above 
two properties are maintained.  It has now been shown \cite{e6} that assuming 
the extended gauge symmetry to be a subgroup of the superstring-inspired 
$E_6$, the success of leptogenesis requires it to be either $SU(3)_C \times 
SU(2)_L \times SU(2)'_R \times U(1)_{Y_L+Y'_R}$ \cite{skew,skew2} or $SU(3)_C 
\times SU(2)_L \times U(1)_Y \times U(1)_N$ \cite{u1n,kema}.  
Only these two gauge groups allow the superfield $N^c$ to have zero quantum
numbers with respect to all of their transformations.  Hence $N^c$ 
may become heavy and decouple from the low-energy phenomenology at the 
supersymmetry-breaking scale.  Remarkably, these 
two subgroups are also the most favored gauge extensions of the SM as 
indicated \cite{favor} by the present neutral-current data from atomic 
parity violation \cite{atom} and precision measurements of the $Z$ width.
Depending on the choice of allowed terms in the superpotential, there are 
two versions of the $U(1)_N$ extension, i.e. Models 1 and 2 of 
Ref.\cite{class}.  Both use the decays of heavy right-handed neutrinos 
(corresponding to the superfield $N^c$) into leptons (or leptoquarks) to 
generate a lepton asymmetry in the early Universe \cite{fuya} 
which gets converted into the present observed baryon asymmetry of the 
Universe through the electroweak sphalerons \cite{krs}. In any other extra 
$U(1)$ model, because its breaking at the TeV scale would introduce $B-L$ 
violating interactions at that scale, the coexistence of the $B+L$ violating 
sphalerons would erase \cite{erase} any lepton or baryon asymmetry that may 
have been created at an earlier epoch of the Universe. 

In this paper we present yet a third alternative which is an elaboration of 
Model 5 of Ref.\cite{class} in the presence of $U(1)_N$.  Here the heavy 
singlet superfield $N^c$ which has $B-L = 1$ is considered to have $B=1$ 
and $L=0$ instead of $B=0$ and $L=-1$ in the usual case.  Since $N^c$ is 
allowed to have a large Majorana mass in the $U(1)_N$ model, its decays 
(into a diquark and a quark as 
well as an antidiquark and an antiquark) may then generate a baryon 
asymmetry of the Universe.  On the other hand, there is no coupling between 
$N^c$ and $\nu$, so there is no canonical seesaw mechanism available for 
$m_\nu$.  As shown previously \cite{skew2,u1n}, there are in general 
3 active and 2 sterile neutrinos in these $E_6$ models. They may acquire 
masses through their mixing with the extra neutral fermions (which are also 
leptons) at the TeV scale.  Hence these neutrino masses are {\it not} 
related to the observed baryon asymmetry of the Universe. Instead, the 
active neutrino masses originate from: $(i)$ the bilinear term $\mu L X^c$ 
where $X^c$ is a new heavy lepton doublet contained in the fundamental 
\underline {27} reprsentation of $E_6$ and which, in contrast to the bilinear 
R-parity breaking models \cite{bil}, conserves R-parity; and $(ii)$ the 
existence of massive sterile neutrinos.  This means that we can also 
accommodate the LSND data \cite{lsnd} if confirmed.  Furthermore, as we 
show below, the decays of the new heavy lepton doublets ($X$ and $X^c$) of 
this model would allow us to map out (partially) the predicted $5 \times 5$ 
neutrino mass matrix.

\section{Neutrino masses in the $U(1)_N$ model of diquark baryogenesis}

The $U(1)_N$ model is defined \cite{e6,u1n} by the charge assignments
\begin{equation}
Q_N = \sqrt {1 \over 40} (6 Y_L + T_{3R} - 9 Y_R),
\end{equation}
whereas the electric charge is given by
\begin{equation}
Q = T_{3L} + Y, ~~~ Y = Y_L + T_{3R} + Y_R,
\end{equation}
under the usual decomposition of $E_6$ into $SU(3)_C \times SU(3)_L \times 
SU(3)_R$.  The various matter superfields belonging to the fundamental 
\underline {27} representation of $E_6$ transform under $SU(3)_C \times 
SU(2)_L \times U(1)_Y \times U(1)_N$ as follows:
\begin{eqnarray}
&& Q = (u,d) \sim (3,2,1/6;1/\sqrt{40}), ~~ u^c \sim (3^*,1,-2/3;
1/\sqrt{40}), ~~ e^c \sim (1,1,1;1/\sqrt{40}), \\ && d^c \sim (3^*,1,1/3;
2/\sqrt{40}), ~~ L = (\nu_e, e) \sim (1,2,-1/2;2/\sqrt{40}), \\ && h \sim 
(3,1,-1/3;-2/\sqrt{40}), ~~ X^c = (E^c,N_E^c) \sim (1,2,1/2;-2/\sqrt{40}), 
\\ && h^c \sim (3^*,1,1/3;-3/\sqrt{40}), ~~ X = (\nu_E,E) \sim (1,2,-1/2;
-3/\sqrt{40}), \\ && S \sim (1,1,0;5/\sqrt{40}), ~~ N^c \sim (1,1,0;0).
\end{eqnarray}

The allowed terms in the superpotential are those which come from the 
decomposition of \underline {27} $\times$ \underline {27} $\times$ \underline 
{27}.  There are eleven such terms.  Five ($Q u^c X^c$, $Q d^c X$, $L e^c X$, 
$S h h^c$, $S X X^c$) are necessary for the usual SM particle masses as well 
as the new heavy particles, i.e. $h$, $h^c$, $X$, $X^c$.  The other six 
($L N^c X^c$, $Q L h^c$, $u^c e^c h$, $d^c N^c h$, $Q Q h$, $u^c d^c h^c$) 
cannot {\it all} be there together because that would induce rapid proton 
decay.  Thus in all $E_6$ models, a discrete symmetry (extension of R-parity) 
has to be imposed to get rid of some of the latter six terms.  In the present 
case, we will adopt a $Z_2 \times Z_2$ discrete symmetry, which can be 
thought of as $(-1)^{3B} \times (-1)^{L}$, i.e. baryon parity and lepton 
parity, which are separately conserved.  This choice is motivated by the 
behavior of the SM baryon and lepton numbers because they are indeed 
separately conserved in that case.

Let us now consider the simplest model, called Model A, resulting from the
$Z_2 \times Z_2$ discrete symmetry.  The first $Z_2=(-1)^{3B}$ is required 
to prevent rapid proton decay and we impose it as follows,
\begin{eqnarray}
Q, u^c, d^c, N^c &:& -1 \\ h, h^c, S, X, X^c, L, e^c &:& +1.
\end{eqnarray}
The allowed trilinear terms in the superpotential are now exactly as in 
Model 5 of Ref.\cite{class}:
\begin{equation}
Q u^c X^c, Q d^c X, L e^c X, S h h^c, S X X^c, d^c N^c h, Q Q h, u^c d^c h^c,
\end{equation}
together with the bilinear terms
\begin{equation}
L X^c, N^c N^c.
\end{equation}
From the above, it is clear that $h$ has $B = -2/3$ (antidiquark) and $h^c$ 
has $B = 2/3$ (diquark). Therefore $N^c$ is  a baryon with  $B = 1$ 
rather than a lepton. 

The second discrete symmetry $Z_2=(-1)^L$ is required to distinguish
between the Higgs and matter supermultiplets. This is exactly analogous
to the minimal supersymmetric standard model in which an appropriate 
R-parity must be imposed for the same reason.
We impose the second $Z_2$ as follows,
\begin{eqnarray}
L, e^c, S_{1,2}, X_{1,2}, X^c_{1,2} &:& -1 \\ 
Q, u^c, d^c, N^c, h, h^c, S_3, X_3, X^c_3 &:& +1.
\end{eqnarray}
The allowed trilinear terms are now further restricted and the complete
superpotential of Model A becomes
\bea
W &=& \lambda_1^{ij}u^c_i Q_j X^c_3 + \lambda_2^{ij}d^c_i Q_j X_3 +
\lambda_3^{ij}e^c_i L_j X_3  + 
\lambda_4^{3ab} S_3  X_a X^c_b +
\lambda_4^{ab3} S_a  X_b X^c_3 +
\lambda_4^{a3b} S_a  X_3 X^c_b +  \nn \\ 
&& \lambda_4^{333} S_3  X_3 X^c_3 +
\lambda_5^{ij} S_3  h_i h^c_j +
\lambda_{6}^{ijk}u^c_i d^c_j h^c_k  + \lambda_7^{ijk}h_i Q_j Q_k + 
\lambda_8^{ijk} d^c_i h_j N^c_k + \nn\\
&& \mu^{ia}L_i X^c_{a} + m_N^{ij} N^c_i N^c_j,
\label{wA}
\eea
where the flavor indices $i,j,k=1,2,3$ run over all 3 flavors while 
$a,b=1,2.$ For further reference we have explicitly written down the structure
of the $\lambda_4$ terms. \Eq{wA} implies that 
$X_3, X^c_3, S_3$ have $L=0$ and are the Higgs superfields 
of this model; but because of the $L X^c_{1,2}$ term, $X^c_{1,2}$ have $L=-1$ 
and $X_{1,2}$ have $L=1$.  On the other hand, because of the $S_{1,2} X_{1,2} 
X^c_3$ and $S_{1,2} X_3 X^c_{1,2}$ terms, $L$ is both $-1$ and $+1$ for 
$S_{1,2}$. In other words, $L$ is not conserved, only $(-1)^L$ is.  Hence 
Majorana neutrino masses are expected in this model because they have even 
$L$ parity.  The model as it stands has no canonical seesaw neutrino mass 
because the $L N^c X^c$ term is absent.  (Hence it also has no provision 
for leptogenesis from the decays of $N$.)  However, because of the 
$\mu L X^c_{1,2}$ term,  4 nonzero masses are still possible 
for the 3 active and 2 sterile neutrinos, as shown below.  

Direct baryogenesis is the distinctive feature of this model. 
Because the heavy Majorana baryon $N^c$ is allowed to have a large 
mass [$Q_N = 0$ for $N^c$ as given in Eq.~(7)], it  may decay 
into the $B-L(=B)=-1$ final states  $\tilde h d^c, $ $ h \tilde{d}^c$ via 
the $\lambda_8$ term in \Eq{wA} as well as into their conjugate states with 
$B-L(=B)=1.$  Technically this mechanism of generating a $B-L$ asymmetry 
is completely analogous to the usual leptogenesis \cite{fuya} except that 
a $B$ asymmetry is created instead of an $L$ asymmetry in the early Universe. 
Since the sphalerons violate $B+L$ but conserve $B-L$, this $B$ asymmetry 
survives as a $B-L$ asymmetry in our present epoch and is observed as a 
baryon asymmetry.  For the details of baryogenesis in $E_6$ models, see 
Ref.\cite{e6}.

Let us now work out the details of how neutrinos become massive in Model A. 
As $U(1)_N$ is broken by the vacuum expectation value of the scalar component 
of $S_3$, the corresponding gauge fermion pairs up with the $S_3$ fermionic 
component to form a massive Dirac particle.  The fermionic components of 
$S_{1,2}$ remain massless and can be considered as sterile neutrinos 
\cite{u1n}.  The $9 \times 9$ mass matrix of the neutral fermions of this 
model with odd $L$ parity, i.e. $\nu_{e, \mu, \tau}, S_{1,2}, \nu_{E_{1,2}}$ 
(from $X_{1,2}$), and $N^c_{E_{1,2}}$ (from $X^c_{1,2}$), is then given by
\begin{equation}
{\cal M} = \left[ \begin{array}{c@{\quad}c@{\quad}c@{\quad}c} 
0 & 0 & 0 & \mu^{ia} \\ 
0 & 0 & \lambda_4^{ab3} v_2 & \lambda_4^{a3b} v_1 \\ 
0 & \lambda_4^{ba3} v_2 & 0 & M_a \delta_{ab} \\ 
(\mu^T)^{ai} & \lambda_4^{b3a}  v_1 & M_a \delta_{ab} & 0 
\end{array} \right],
\label{9x9}
\end{equation}
where $v_1 = \langle \tilde \nu_{E_3} \rangle$, $v_2 = \langle \tilde 
N^c_{E_3} \rangle$, and $M_{1,2}$ are the mass eigenvalues of the 
$X_{1,2} X^c_{1,2}$ 
mass matrix, proportional to $\langle \tilde S_3 \rangle$.  It is clear 
from the above that 4 of the 9 fields are heavy with masses $M_1, M_1, 
M_2, M_2$ approximately, and that 4 others are light with seesaw masses 
inversely proportional to $M_{1,2}$. One neutrino, however, 
remains massless in this model.  The $5 \times 5$ 
reduced mass matrix spanning the 3 active and 2 sterile neutrinos thus
becomes
\begin{equation}
{\cal M}_\nu = \left[ \begin{array} {c@{\quad}c} 
0 & \sum_c \mu^{i c} \lambda_4^{cb3} v_2 M_c^{-1} \\ 
\sum_c \lambda_4^{ac3}  (\mu^T)^{cj} v_2 M_c^{-1} & 
\sum_c ( \lambda_4^{ac3} \lambda_4^{c3b} + \lambda_4^{a3c}\lambda_4^{cb3}) 
v_1 v_2 M_c^{-1} \end{array} \right].
\label{5x5}
\end{equation}
This shows explicitly that without the sterile neutrinos and without the
bilinear term $\mu L X^c$, the active 
neutrinos themselves would be massless.  Such a result has also been obtained 
recently in a very different model \cite{mrs} of decaying sterile neutrinos 
from large extra dimensions.  To obtain realistic neutrino masses, we note 
that the sterile neutrino masses may be of order 1 eV, and the off-diagonal 
entries in \Eq{5x5} of order 0.1 to 0.01 eV, then the active neutrino masses 
may be of order $10^{-2}$ to $10^{-4}$ eV.

\section{Related phenomenology at colliders}

We start our discussion regarding the structure of the neutrino mass 
matrix of \Eq{9x9} with three comments.  First, the new heavy supefields 
$N^c_{1,2,3}$ do not contribute to \Eq{9x9}. Thus the observed baryon 
asymmetry of the Universe, created via the $\lambda_8$ terms of \Eq{wA},
is not related in any way to the measured neutrino masses.  Secondly, the 
mass matrix of \Eq{9x9} involves only {\it leptons}; no superpartners such 
as gauginos and Higgsinos are there.  Thus, in spite of the presence of 
the $\mu$ entries from the bilinear $LX^c$ term, R-parity is conserved here. 
Third, the Yukawa coupling matrices $\lambda_4$ should be nonvanishing. The 
matrix $\lambda_4^{3ab}$ which gives masses to the new leptons $X,X^c$ can 
be chosen to be diagonal without loss of generality.  The matrices 
$\lambda_4^{ab3}$ and $\lambda_4^{a3b}$ provide the mass terms $S\nu_E$ 
and $S N_E^c,$ respectively, which must also be nonzero.  Otherwise, all 
neutrinos would be massless as seen from \Eq{5x5}.

Because the neutrino masses come entirely from new physics at the TeV 
scale, our model has a good chance of getting tested at future collider 
experiments.  The first issue to be determined is the existence
of the sterile neutrinos.   As $S_{1,2}$ have gauge couplings only to 
$Z',$ the invisible width of $Z'$ is predicted to have the property
\begin{equation}
\Gamma (Z' \to \nu \bar \nu + S \bar S) = \left( {62 \over 15} \right) 
\Gamma (Z' \to l^- l^+),
\end{equation}
which should distinguish it from other $Z'$ models.  Also, neutrino 
oscillations between the 3 active and 2 sterile neutrinos are possible 
and {\it natural} in our model, so the LSND data can be accommodated. 
But even if the LSND results turn out to be erroneous, it is still possible 
that future long-baseline experiments and neutrino factories may see the 
conversion to sterile neutrinos at some different $\Delta m^2$.

Another prediction of our model is the existence of the two heavy
lepton doublets $X=(\nu_E,E)$ and $X^c=(E^c,N_E^c)$ which are
almost degenerate in mass. 
Once produced, these new heavy particles may decay via gauge and Yukawa 
interactions. Because of the $S_{1,2}$ mixings with the heavy neutral leptons
$\nu_E, N^c_E$, the gauge interactions induce the decays
\bea
&& E_{1,2} \to W^- S_{1,2}\,, \qquad E^c_{1,2} \to W^+ S_{1,2}\,, \nn\\
&& \nu_{E_{1,2}} \to Z\;S_{1,2}\,,\qquad N^c_{E_{1,2}} \to Z\; S_{1,2} \,.
\eea
In principle, these decay processes measure directly the corresponding
neutrino mixing angles in \Eq{9x9} which are predicted to be 
of order $\sim \mu/M$. On the other hand, the Yukawa couplings $\lambda_4$ 
in \Eq{wA} give rise directly to the competing decays
\bea
&& E_{1,2} \to H^- S_{1,2}  \,, \qquad E^c_{1,2} \to H^+ S_{1,2}\,, \nn\\ 
&& \nu_{E_{1,2}} \to H^0 S_{1,2}  \,, \qquad N^c_{E_{1,2}} \to H^0 S_{1,2} \,,
\label{d1}
\eea
where the physical charged Higgs boson $H^+$($H^0$) is a mixture of 
$\tilde E_3$($\tilde \nu_{E_3}$) and $\tilde E^c_3$($\tilde N^c_{E_3}$). 
In our model, to get the correct order of magnitude for the neutrino masses, 
we estimate $\mu/M\sim 10^{-6}$ while $\lambda_4\sim 10^{-5}.$ 
Therefore the latter decays should be dominant.  
Notice however that because the different sterile neutrino final states
cannot be distinguished from each other, the structure of $\lambda_4$ 
couplings cannot be determined from these decays. 

From the experimental point of view, decay modes of $X,X^c$ with charged 
leptons in the final state offer much better signals for testing our model.
Because the bilinear term $LX^c$ mixes the known charged leptons with the 
new heavy charged leptons, the decays
\bea
&& N^c_{E_a} \to W^- e^c_i \,, \qquad
E^c_a \to Z e^c_i \,,
\label{d2}
\eea
should occur.  These are proportional to $\mu^{ia}/M_a.$   Hence the 
ratio of the branching fractions of the decays \rfn{d2} over \rfn{d1} is
predicted to be $(\mu^{ia}/(\sum\lambda_4)M)^2\sim 10^{-2}.$ 
If 1\% precision will be achieved in determining these branching 
fractions, one should then be able to obtain the structure of $\mu^{ia}$ 
for all $i=e,\mu,\tau$ and $a=1,2$.  In this respect, our proposal is similar 
to that of Ref.\cite{bil}.

Let us assume now that all $\lambda_4$ couplings are diagonal and 
equal. In that case, the measured $\mu^{ia}/\sum\lambda_4$ 
determines the flavor structure of the neutrino masses according to 
the $5\times 5$ neutrino mass matrix of \Eq{5x5}.
This might be a crude approximation but it allows us to test our 
neutrino mass matrix up to an overall scale, which must then 
be determined from other neutrino experiments.

We finish this Section with a comment on the hadron-collider phenomenology.
Our model necessarily predicts the existence of diquark superfields 
$h,h^c$ with masses of order $M_{Z'}$.  Perhaps the most distinctive 
experimental signatures are then s-channel diquark $\tilde h$ and 
$\tilde h^c$ resonances at hadron colliders.  At the LHC, the initial state 
from 2 valence quarks carries $B=2/3$, hence a diquark resonance may occur 
without suppression.  This allows us to test the existence of $\tilde h$ 
and $\tilde h^c$ to about 5 TeV \cite{res}.

\section{Model B}

An important variation of our basic model, called Model B,  
can be obtained by choosing $N^c_{1,2}$ to have $(-1,+1)$ under $Z_2 
\times Z_2$ as in Model A, but $N^c_3$ is assigned $(+1,-1)$ instead. 
The decays of $N^c_{1,2}$ will again generate a baryon asymmetry. [Note that 
the condition of CP violation requires at least two such heavy superfields.] 
But now $N^c_3$ is a lepton and the allowed term $L N^c_3 X^c_3$ enables 
one linear combination of 
$\nu_e$, $\nu_\mu$, and $\nu_\tau$ to acquire a canonical seesaw mass.  This 
is an excellent opportunity for choosing $\nu_\mu \cos \theta + \nu_\tau \sin 
\theta$ to be massive with $\theta$ near $\pi/4$ for maximal mixing to 
explain the atmospheric neutrino data.  The other 2 active neutrinos will 
both become massive from \Eq{5x5} as before.  This may provide a rationale 
for having one active neutrino mass much larger than the other two, and 
allow all 3 active neutrinos to be massive instead of only 2 as in Model A.

As an illustration, let
\begin{equation}
\left( \begin{array} {c} \nu_1 \\ \nu_2 \\ \nu_3 \end{array} \right) = 
\left( \begin{array} {c@{\quad}c@{\quad}c} 1/\sqrt 2 & -1/2 & 1/2 \\ 
1/\sqrt 2 & 1/2 & -1/2 \\ 0 & 1/\sqrt 2 & 1/\sqrt 2 \end{array} \right) 
\left( \begin{array} {c} \nu_e \\ \nu_\mu \\ \nu_\tau \end{array} \right),
\end{equation}
and consider the following $5 \times 5$ neutrino mass matrix in the basis 
$(\nu_{1,2,3}, S_{1,2})$:
\begin{equation}
{\cal M}_\nu = \left[ \begin{array} {c@{\quad}c@{\quad}c@{\quad}c@{\quad}c} 
0 & 0 & 0 & 0 & 0 \\ 0 & 0 & 0 & \mu_1 & \mu_2 \\ 0 & 0 & m_3 & 0 & 0 \\ 
0 & \mu_1 & 0 & 0 & M \\ 0 & \mu_2 & 0 & M & 0 \end{array} \right].
\end{equation}
Let $m_3 \sim 0.05$ eV and choose
\begin{equation}
m_2 \sim {2 \mu_1 \mu_2 \over M} \sim 3 \times 10^{-3} ~{\rm eV},
\end{equation}
then we have maximal atmospheric neutrino oscillations with $\Delta m^2 \sim 
2.5 \times 10^{-3}$ eV$^2$ and maximal solar neutrino oscillations with 
$\Delta m^2 \sim 10^{-5}$ eV$^2$.  If LSND data are also to be explained, 
then we can set $M \sim 2$ eV, $\mu_1 \sim 0.5$ eV, $\mu_2 \sim 6 \times 
10^{-3}$ eV, so that
\begin{equation}
P_{\mu e} \sim {1 \over 4} \left( {\mu_1^2 + \mu_2^2 \over M^2} \right)^2 
\sim 10^{-3},
\end{equation}
in agreement with experiment.

\section{Conclusions}

In the context of superstring-inspired $E_6$ extensions of the 
supersymmetric SM, we have shown how the three active and two 
sterile neutrinos obtain realistic masses in two 
$SU(3)_C \times SU(2)_L \times U(1)_Y \times U(1)_N$ models of diquark 
baryogenesis. This extra $ U(1)_N$ is favored over other
extra $U(1)$ models by recent neutral-current data from atomic parity 
violation and the invisible $Z$ width.
In our scenario, neutrino masses originate from new physics at the TeV 
scale and are not related to the baryogenesis parameters at a very high 
scale. The active neutrinos acquire masses from the R-parity conserving
bilinear superpotential term $\mu L X^c$ and nonzero sterile neutrino 
masses.  Some of the neutrino mass matrix entries can be directly tested
with the decays of the new heavy lepton doublets $X$ and $X^c$ at future 
collider experiments.  We also predict the fixed properties of the extra 
$Z'$ boson, the necessary existence of two sterile neutrinos, and the 
observation of diquark resonances at the LHC or Tevatron.

{\it Acknowledgement.} 
This work was supported in part by the U.~S.~Department of Energy under Grant 
No.~DE-FG03-94ER40837.

\newpage
\bibliographystyle{unsrt}

\end{document}